\documentclass[useAMS,usenatbib,usegraphicx]{mn2e}

\title[Have proto-planetary discs formed planets?]{Have 
proto-planetary discs formed planets?}
\author[J. S. Greaves et al.]{J. S. Greaves$^{1}$\thanks{E-mail: jsg5, at 
st-andrews.ac.uk (JSG).} and W. K. M. Rice$^{2}$\\
$^{1}$SUPA, Physics and Astronomy, St Andrews University, North Haugh, 
St Andrews, Fife 
KY16, U.K. \\
$^{2}$SUPA, Institute for Astronomy, Royal Observatory, Blackford Hill, 
Edinburgh EH9 3HJ 
}

\begin{document}

\date{Accepted 2010. Received 2010; in original form 2010}

\pagerange{\pageref{firstpage}--\pageref{lastpage}} \pubyear{2009}

\maketitle

\label{firstpage}

\begin{abstract}

It has recently been noted that many discs around T Tauri stars appear to 
comprise only a few Jupiter-masses of gas and dust. Using millimetre 
surveys of discs within six local star-formation regions, we confirm this 
result, and find that only a few percent of young stars have enough 
circumstellar material to build gas giant planets, in standard core 
accretion models. Since the frequency of observed exo-planets is greater 
than this, there is a `missing mass' problem. As alternatives to simply 
adjusting the conversion of dust-flux to disc mass, we investigate three 
other classes of solution. Migration of planets could hypothetically sweep 
up the disc mass reservoir more efficiently, but trends in multi-planet 
systems do not support such a model, and theoretical models suggest that 
the gas accretion timescale is too short for migration to sweep the disc. 
Enhanced inner-disc mass reservoirs are possible, agreeing with 
predictions of disc evolution through self-gravity, but not adding to 
millimetre dust-flux as the inner disc is optically thick. Finally, the 
incidence of massive discs is shown to be higher at the {\it proto}stellar 
stages, Classes 0 and I, where discs substantial enough to form planets 
via core accretion are abundant enough to match the frequency of 
exo-planets. Gravitational instability may also operate in the Class~0 
epoch, where half the objects have potentially unstable discs of 
$\ga$30~\% of the stellar mass. However, recent calculations indicate that 
forming gas giants inside 50~AU by instability is unlikely, even in such 
massive discs. Overall, the results presented suggest that the canonically 
`proto-planetary' discs of Class~II T Tauri stars {\bf have globally low 
masses in dust observable at millimetre wavelengths, and conversion to 
larger bodies (anywhere from small rocks up to planetary cores) must 
already have occurred.} 

\end{abstract}


\section{Introduction}

It is often assumed that the discs of classical T Tauri stars are those in 
which planets form. These `Class II' objects have disc-dominated dust 
emission in the infrared to millimetre, in contrast to fainter remnant 
discs around Class III (weak-line) T Tauri stars, or the 
envelope-dominated emission of the earlier-stage Class 0 and I protostars. 
Provided that dust persists for a few Myr in Class~II discs, along with 
the gas component that contains the bulk of the mass \citep{dent}, then 
gas giants should be able to form by core accretion followed by 
accumulation of a thick atmosphere \citep{pollack,hubickyj}. However, it 
has been noted recently that the masses of T~Tauri discs as derived from 
millimetre observations are typically small compared to the reservoirs 
needed for gas-giant formation. \citet{aw07} inferred that the median 
Class II disc comprises only 5 Jupiter masses of dust and gas, from deep 
surveys of the Taurus and Ophiuchus regions. In contrast, at least 20 
Jupiter masses was present in the Minimum Mass Solar Nebula (MMSN), on the 
basis of the rocky content now locked up in planetary cores \citep{davis}. 
In models, a mass-reservoir a few times larger is actually needed for 
core-accretion to successfully build the the Sun's giant planets. For 
example, \citet{dodson} find that a 120 Jupiter-mass\footnote{The 
convenient relation $M_{Jupiter)} \approx M_{Sun}/1000$ is adopted 
throughout, rather than the true inverse mass ratio of 1047.} or 6-MMSN 
disc can make all four giants. There are also examples of more massive 
exo-planets than the Sun's giants, probably requiring more substantial 
discs from which to form. Hence, if exo-discs only comprise a few 
Jupiter-masses, there is a `missing mass' problem at these early stages.

Further, the surface densities that are the critical parameter in core 
accretion models are typically measured to be lower than expected, for 
example in relation to forming Jupiter at around 5~AU \citep{aw07b}. 
Millimetre dust emission is a good choice for measuring total disc-masses, 
being of low optical depth and reasonably bright, but limited angular 
resolution means that material at a few AU is not resolved. Although the 
bulk of the disc mass should be at large scales \citep{wyatt}, 
discontinuous profiles with a strong central mass peak can not be ruled 
out. However, if the profiles are smooth, the typical outer radii of 
$\approx 200$~AU \citep{aw07b} are much larger than the Solar System, 
exacerbating the missing-mass problem by further reducing the 
inferred inner-disc surface densities.

Several types of solution to the missing-mass problem have been proposed. 
The conversion of millimetre flux to dust mass is via the grain 
temperature and emissivity, and while the former can be estimated from the 
wavelength of peak emission, the latter is more uncertain and comes from 
modelling of grain properties \citep{draine}. Decreasing the adopted 
emissivity would enhance the derived disc masses, and this could be 
reasonable for a different composition, or a size distribution where much 
of the mass is in large grains. Such `pebbles' are not conspicuous in the 
millimetre measurements but can be detected by radio continuum emission 
\citep{rodmann}. In addition, the total disc masses could be wrong if the 
assumed gas-to-dust mass ratio of 100, i.e. inherited from the 
interstellar medium, is in error. A higher conversion factor would be 
correct if some of the dust has evolved into non-observable forms ({\bf 
anywhere from small pebbles up to} planetesimals) but the gas is still all 
in the disc. The true disc masses should then be higher if a more 
appropriate gas-to-dust ratio were used. Similarly, if some of the gas has 
also been accreted into planets, the disc mass estimates are even more 
uncertain but will underestimate the initial reservoir. {\bf Generally,  
\citet{aw07} have shown that the disc masses estimated via the 
amount of accretion onto the star are substantially higher than the masses 
derived directly from the observed dust, confirming that 
there is indeed a discrepancy within the available data.}

\begin{table*}
 \caption{Summary of disc masses from millimetre surveys. Cluster star 
counts are from \citet{porras} except for Lup \citep{merin} and Tau/Aur 
\citep{kenyon}, and may not be identical regions to those surveyed for 
discs. Ages given are the medians within each cluster from \citet {palla}. 
Data at $\lambda$ mm are from interferometers for ONC and IC~348 and 
otherwise from single dishes -- these photometry results may include cloud 
emission in crowded regions. The number of discs detected is given as a 
subset of the total number of targets including all spectral types and 
evolutionary stages; in IC~348 a net detection only was made over all the 
stars observed. Subsequent columns list the fraction of MMSN discs (above 
20~M$_{Jup}$) and Poisson errors, the median mass (approximate where over 
half the data are upper limits), and the 100~\%-complete mass limit of the 
survey.
}
 \label{tab:clusters}
 \begin{tabular}{@{}lcrcccccl}
  \hline
cluster	& age	& N$_{stars}$ & $\lambda$ & N$_{observed}$ & $f_{MMSN}$	& $M_{disc}$ 		& $M_{complete}$& reference 	\\
	& (Myr)	&	      & (mm)	  & ($N_{discs}$)  & (\%)	& (median, $M_{Jup}$) 	& ($M_{Jup}$)	&		\\
ONC	& 1	& 1471	& 0.88		& 55 (26)	& 11 $\pm$ 4	& $\la 8$	& 8	& \citet{mann} 		\\
Oph	& 1	& 337	& 0.35-1.3	& 147 (64)	& 18 $\pm$ 4	& $\la 6$	& 10	& \citet{aw07}		\\
Tau/Aur & 1.5	& 289	& 0.35-1.3	& 153 (94)	& 17 $\pm$ 3	& 1		& 1	& \citet{aw05}		\\
IC 348	& 1.5	& 345	& 3		& 95 (0)	& $\sim 0$	& 2		& 25	& \citet{carpenter}	\\
Lup	& 2	& 159	& 1.3		& 32 (12)	& 6 $\pm$ 4	& $\la 5$	& 7	& \citet{nurnberger}	\\
Cha	& 2	& 180	& 1.3		& 36 (16)	& 8 $\pm$ 5	& $\la 7$	& 16	& \citet{henning}	\\
  \hline
 \end{tabular}
\end{table*}

There are however some difficulties with these kinds of solutions to the 
missing-mass problem. Firstly, a bulk increase in the adopted emissivity 
will affect all systems, and the few discs of $\ga 100 M_{Jupiter}$ 
\citep{aw05,aw07,aw07b} would then be comparably massive to the host star. 
This situation is likely to be dynamically unstable, i.e. such a disc 
would tend to fragment to form a companion star or brown dwarf, and not 
remain to be observed in the Class~II era. Secondly, where larger grains 
are detected in long-wavelength data, the derived disc masses do not 
always increase. \citet{rodmann} found systems where the millimetre 
spectral index is small -- a signature of grain growth \citep{draine} -- 
with three discs having indices as low as 0.5-0.8 in the 1-7~mm regime. 
However, masses derived from their 7~mm fluxes range from 20~\% lower to 
70~\% higher than the estimates based on 0.85~mm data, rather than being 
substantially increased. Thirdly, the bulk H$_2$ gas component has in some 
cases been detected by infrared spectroscopy, although this is only 
sensitive to warm layers and so the total gas reservoir must again be 
estimated. {\bf In their models} \citet{bary} find that only a few Jupiter 
masses of hydrogen at $\sim 10-30$~AU is needed to reproduce line fluxes, 
but many systems a few Myr old are not detected, so predominantly high 
gas-to-dust mass ratios are not supported. The gas could already have been 
accreted into planetary atmsopheres in favourable cases \citep{hubickyj}, 
but it may instead have been simply lost by mechanisms such as 
photo-evaporation \citep{ercolano}.

Here we examine some alternative solutions to the missing-mass problem. We 
first compile a millimetre-based distribution of $M_{disc}$ over several 
regions of young stars, and consider if there are globally enough 
high-mass discs to match the fraction of Sun-like stars observed to host 
gas giants. Finding that there is still a disc-mass deficit, we then 
examine three alternative classes of solution: that the mass reservoir 
could be used more efficiently if planets migrate while forming; that mass 
is hidden at small radii because it is optically thick; and that the mass 
has disappeared because it is accreted into planets very early on, even at 
the protostellar stage.

\section{Disc material}

Star formation near the Sun is well characterised by the molecular clouds 
of the Gould Belt, lying at distances of around 150 to 500~pc 
\citep{gould}. The belt includes at least a dozen regions of young stars, 
of which six have deep millimetre-wavelength surveys for dust discs. Other 
Gould Belt regions have few millimetre-detected discs, either because the 
stars are still mostly very deeply embedded so that intensive millimetre 
interfometry is needed to identify compact structures (e.g. Serpens, 
Corona Australis, IC 5146), or the large cloud areas have yet to be 
thoroughly mapped (e.g. Cepheus Flare, Pipe Nebula). The disc surveys 
included here are listed in Table~1, and cover the Taurus/Auriga, 
$\rho$~Ophiuchus, Chamaeleon and Lupus regions, the Orion Nebula Cluster 
and the IC~348 cluster in Perseus. The median stellar ages in these clouds 
\citep{palla} are 1-2~Myr, although the disc surveys include Class~O and I 
protostars that may be $\ll 1$~Myr old. The surveys have target selection 
biases, and in particular faint Class III systems may be 
under-represented. Also, not all the young stars were observed, with 
surveys sampling $\sim 30-150$ targets out of $\sim 150-1500+$ objects in 
each cluster. However, these data are the best snapshot presently 
available of the properties of potential planet-forming discs. The 
environments range from sparse star formation in Tau/Aur to the dense ONC, 
with the total star-count greatly dominated by this latter region; hence, 
its discs are of particular interest as they may represent archetypal 
planet-forming conditions.

Subsequent discussion implicitly assumes that these discs we observe today 
are equivalent to those that formed planets in the past -- typically a few 
Gyr ago when considering the Sun-like hosts of observed exo-planets. We 
also assume that all the star-formation environments that existed in the 
past are also observed today; that the exo-planets observed are not from a 
peculiar subset of stars; and that the mode of planet formation that 
dominated in the past has not greatly altered as for example the Galaxy 
has become more enriched in heavy elements.

\subsection{Millimetre-based disc masses}

Table~1 summarises the results of the these millimetre disc surveys, and 
presents global properties for all the systems observed in each region. 
The results are adopted directly from the survey papers without attempting 
to align slightly different assumptions about dust temperatures and 
emissivities. All the surveys were complete down to an MMSN of 
20~$M_{Jup}$, apart from that of IC~348 where this represents a $2.5 
\sigma$ upper limit for each star.

With this very comprehensive dataset of 518 stars, it is clear that the 
fraction hosting an MMSN when observed is small. No cluster has an 
incidence $f_{MMSN} > 20$~\% and the overall average is $f_{MMSN} = 12 \pm 
2$~\% (Poisson error). The median disc of each cluster comprises only a 
few Jupiter masses, down to as little as $1-2 M_{Jup}$ in the deepest 
data. There are no marked differences between sparse and dense clusters, 
and $f_{MMSN}$ does not decline significantly with time. All of the 
clusters are within $\approx 1.5\sigma$ of the global $f_{MMSN}$, except 
for IC~348 where no MMSN are seen (but the survey is barely sensitive to 
this depth).

Figure~1 shows the estimated underlying distribution of disc masses. The 
discs have been divided among bins covering a factor of 3 in mass, with 
the highest three bins representing an MMSN or more. The raw counts were 
scaled up to infer the numbers of discs that would be expected for all the 
stars known in the cluster (Table~1), and these counts are then stacked to 
show the estimated contribution of each cluster. For uniformity the upper 
limits were treated as values, so the plotted distribution is the {\it 
most optimistic} one from the view of planet-forming potential. Two 
clusters were treated as special cases: the ONC, where completeness 
information \citep{mann} was used to infer limits, and IC~348, where the 
nominal measured fluxes \citep{carpenter} were treated as data points 
where positive, and the mean 2~$M_{Jup}$ disc was adopted for all other 
stars\footnote{This avoids the problem of logarithms of negative values 
but is not strictly flux-conserving, with the mean signal per disc being 
inflated to the $+4\sigma$ bound.}. For Tau/Aur and Oph only, some limits 
or detections below 2~$M_{Jup}$ were reached, but these are omitted from 
the plot for clarity.

Figure~1 also implies that the number of discs with MMSN-like planet 
forming potential is low. Including the scalings by total number of stars 
per cluster, $f_{MMSN}$ is now 11~\%, very similar to the uncorrected 
value above of 12~\%. The proportion of discs in the top two bins, above 
the thresholds in many gas giant formation models, is only about 0.5~\%, 
and such massive discs were only found in the Tau/Aur and Oph regions.

\begin{figure}
\label{fig1}
\includegraphics[width=57mm,height=68mm,angle=270]{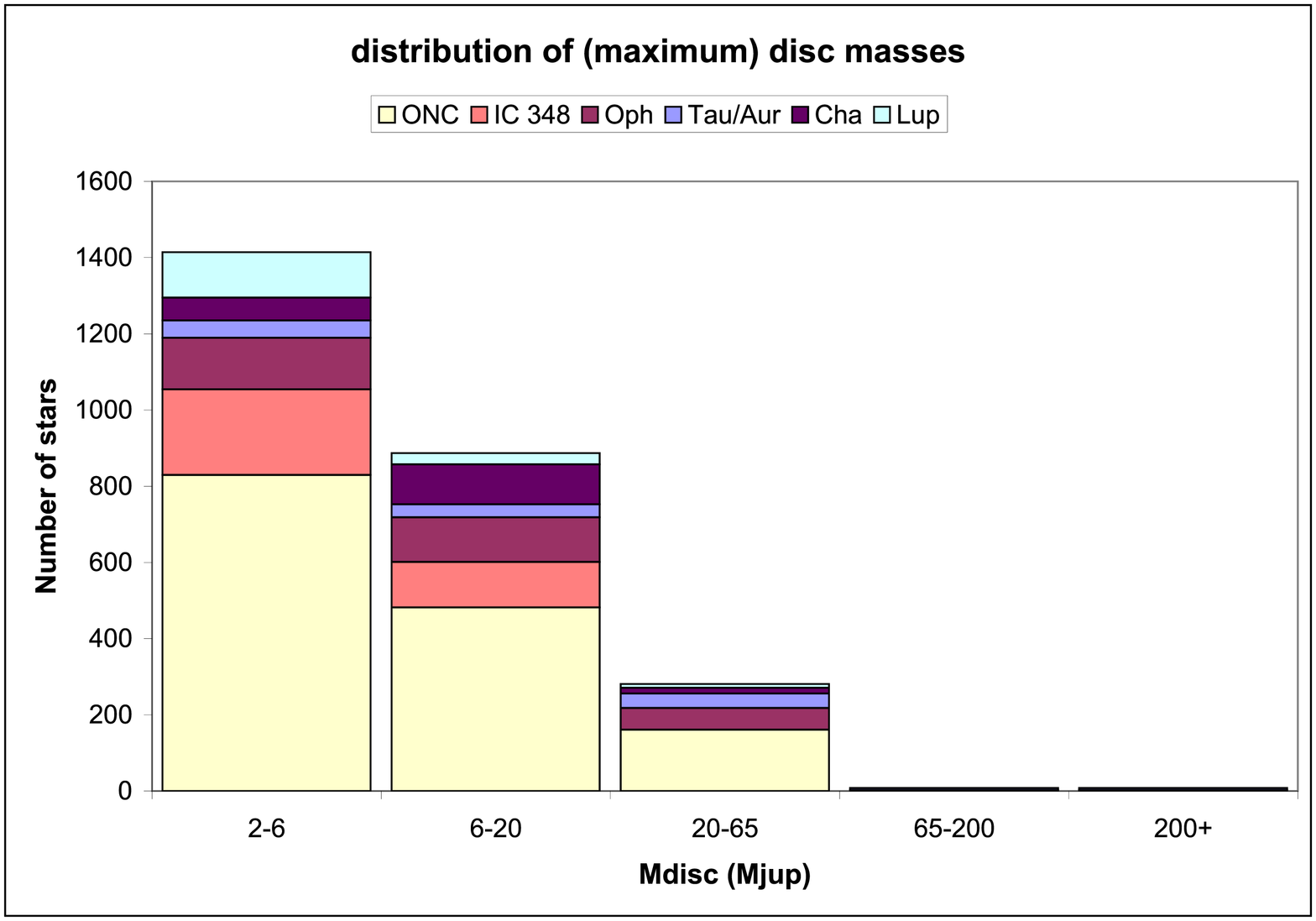}
\vspace*{12mm}
\caption{Distribution of disc masses from surveys listed in Table~1, 
scaled by the number of stars in each cluster and treating upper limits 
as values, hence maximising the $M_{disc}$ bin they lie in (see text). The 
three right-most bins correspond to one MMSN upwards. On the left side, 
sparse data for low-mass discs (7~\% of stars) are omitted for clarity.
}
\end{figure}

\begin{figure*}
\label{fig2}
\vspace*{10mm}
\includegraphics[width=55mm,height=80mm,angle=270]{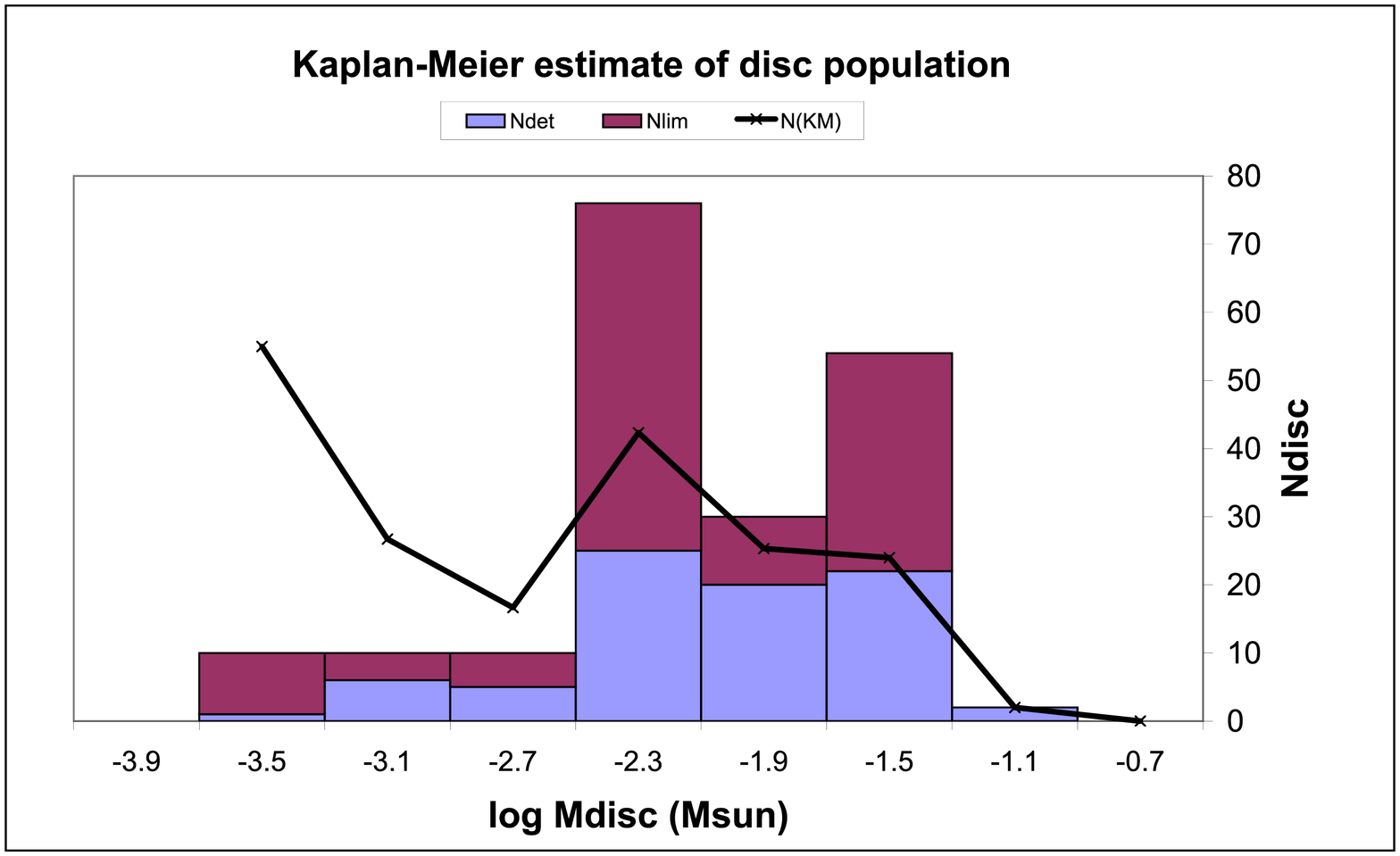}
\includegraphics[width=55mm,height=80mm,angle=270]{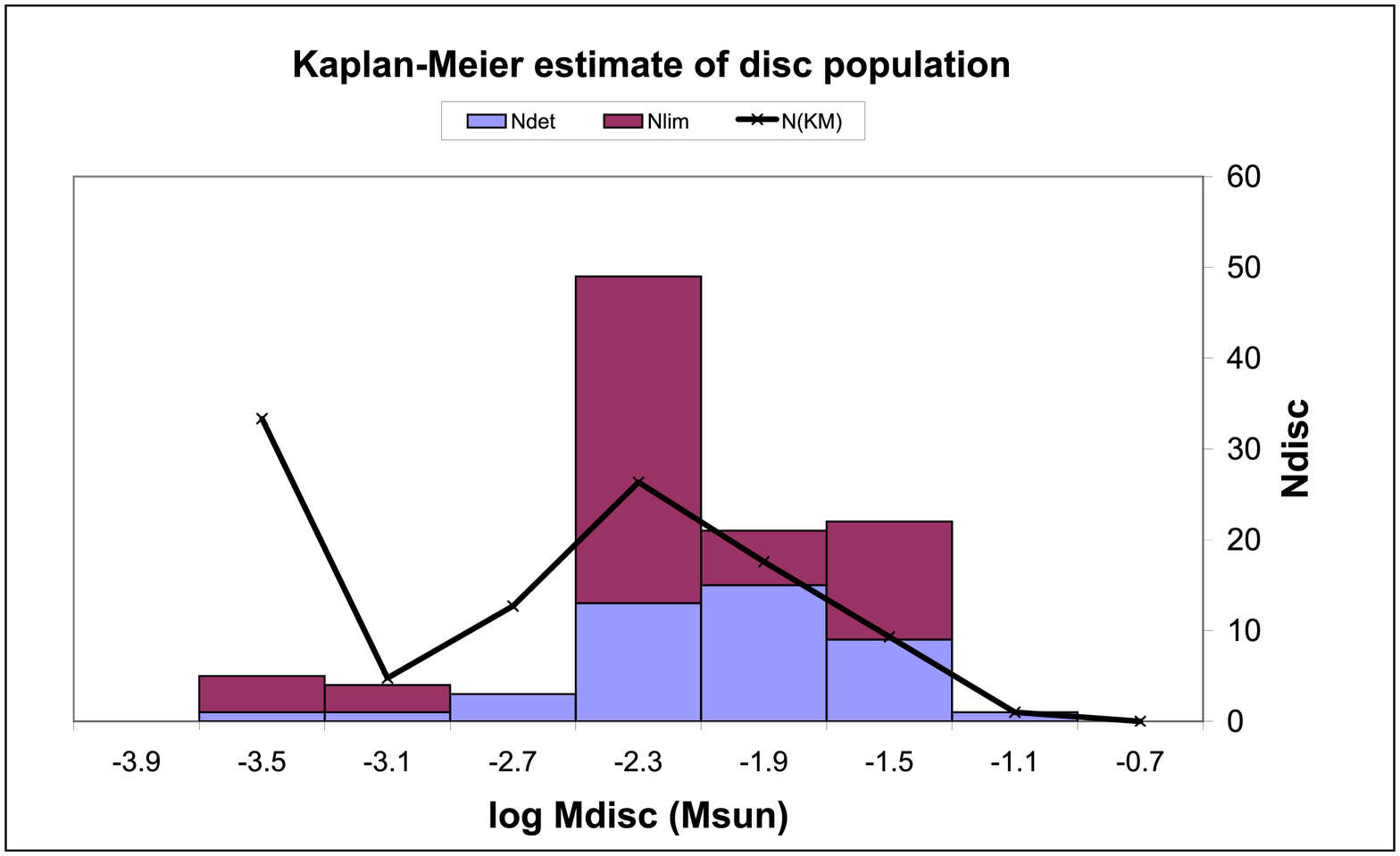}
\caption{Kaplan-Meier estimates of the underlying disc population (see 
text). The upper and lower stacked histogram bars indicate limits and 
detections in each mass bin respectively, and the line is the K-M fit. 
The right-most three bins comprise 1 MMSN upwards. Left: weighting each 
cluster equally; Right: weighting each cluster proportional to its total 
number of stars from Table~1.
}
\end{figure*}

\subsection{Disc population}

Survival analysis is used next to make a better estimate of the underlying 
distribution of disc masses, given the varying limits between and within 
surveys. The Kaplan-Meier estimator is appropriate when the limit obtained 
is independent of the mass of the disc being observed, a condition that is 
reasonably well met here. The K-M tests were performed on the 
log-disc-mass data using the ASURV package \citep{lavalley} and the 
results are shown in Figure~2. In the first panel, all of the 6 clusters 
of Table~1 were weighted equally, by taking 32 stars from each (i.e. all 
of those observed in Lupus and a random set of 32 from each of the other 
clusters). In the second panel, the clusters were weighted according to 
the total number of stars in each, i.e. taking all 55 observed discs from 
the ONC and a proportionally smaller random set from the other clusters 
(down to 6 stars in Lupus, i.e. $55 \times (N_{Lup}=159) / 
(N_{ONC}=1471)$). These two approaches attempt to take into account a 
different mix of star-forming environments that could have occured at past 
times.

In both cases, Figure~2 shows a broad distribution of masses plus a large 
tail of very insubstantial discs. The main distribution peaks at 5 Jupiter 
masses, while about 30~\% of stars have such negligible discs (left-most 
bin) that the dust mass is $\la 1 M_{\oplus}$ -- these might not even be 
able to make a single terrestrial planet. At the high-mass end, weighting 
all the clusters equally yields an MMSN fraction of 13.5~\%, while the 
$N_{star}$-weighted calculation gives 10~\% MMSN. These results are very 
similar to the previous results of $f_{MMSN} \approx 11-12$~\%. Hence, 
even if the modes of star formation differed in the past, the fraction of 
discs capable of forming gas gaint planets appears to have been low, and 
many stars seem to have been incapable of any planet formation.

\subsection{Disc mass evolution}

It has already been suggested that higher disc masses at earlier stages 
might be more conducive to planet formation, and for example \citet{aw07} 
have shown that this is the case for Taurus and Ophiuchus. There the 
median disc mass is 5~$M_{Jup}$ for Class~II stars but 15~$M_{Jup}$ for 
the precursor Class~I objects. Table~2 gives a similar summary of the data 
for all six star-forming regions of Table~1. No weighting of the clusters 
is made, as wide variations in detection-rates and proportions of 
different classes observed make this difficult to do systematically. Where 
objects are not explicitly classified, we treat as all-Class~II both the 
ONC proplyds (following \citet{mann}) and the slightly older IC~348 stars 
{\bf (for consistency, although in fact the proportion of II:III objects 
here is about 30:70} \citep{muench}). Class~III stars are not included in 
Table~2 because there are very few dust detections and so average 
properties are not meaningful.

\begin{table}
 \caption{Disc masses sub-divided by object class. Data from all 6 regions 
of Table~1 are added together without weighting. There are 38, 25 and 339 
objects in Classes I, FS and II respectively, and the corresponding 
fractions of upper-limit data are 5, 24 and 43~\%. For each pair of lines,  
the upper includes detections only, and the lower (bracketed) includes 
the upper-limit systems but treated as zero disc-mass. The Poisson error 
is given for the frequency of MMSN. 
}
 \label{tab:classes}
 \begin{tabular}{@{}lccccc}
  \hline
		& Class I	& Flat Spectrum	& Class II \\
  \hline
mean $M_{disc}$ &  41		& 14		& 14 	\\
($M_{Jup}$)     &  (39)		& (11)		& (8) 	\\
median $M_{disc}$& $\approx15$	& 10		& 9	\\
($M_{Jup}$)    	& ($\approx15$)& (10)		& (2)	\\
$f_{MMSN}$	&  47 $\pm$ 11	& 26 $\pm$ 12	& 22 $\pm$ 3 \\
(\%)		& (45 $\pm$ 11)	& (20 $\pm$ 9)	& (13 $\pm$ 2) \\
  \hline
 \end{tabular}
\end{table}

Table~2 confirms that the fraction of MMSN is higher for Class~I objects 
than at either of the subsequent Flat-Spectrum or Class~II stages. The 
median disc mass is below an MMSN at all stages, but at Class~I only just 
so\footnote{Millimetre photometry may somewhat over-estimate Class~I disc 
masses, if the remnant envelopes contribute some flux.}, and also 7/38 
objects (18~$\pm$~7~\%) have disc-masses of 2-25~MMSN as in many 
giant-planet formation models. The previous results on the low fraction of 
MMSN are clearly dominated by the preponderence of Class~II discs, with 
Class~I/FS objects representing briefer stages and thus fewer examples 
seen at snapshots in time. The mean disc-mass declines after Class~I, but 
the median mass only drops after the FS stage, when a large tail of upper 
limits appears within Class~II. Combined with the remnant discs of 
Class~III, many systems have very low masses by the few-Myr epoch.

\subsection{Comparison to planetary systems}

We now consider the mass budget needed in early-time discs to form the 
observed extrasolar planets.  Analysis of Doppler surveys \citep{cumming} 
shows that approximately 18~\% of Sun-like stars should host a giant 
planet, of above about Saturn's mass and orbiting within 20~AU. The 
results are based on extrapolations of detections made so far, but are 
consistent with different extrapolation methods. The typical planet mass 
is about that of Jupiter, and a little over half of these systems should 
host a body orbiting at about 3~AU outwards, comparable to the location of 
the Sun's gas giants. Within this region, microlensing surveys indicate 
that Neptune-mass planets are at least three times more abundant than 
Jupiters \citep{sumi}. Multiple planetary systems are expected to be 
common, with over a quarter of those found by Doppler wobble already known 
to have two or more gas giants \citep{wright}.

If giant planets orbiting at a few AU are common, early-time disc-mass 
reservoirs of a few MMSN are expected. A high surface density of solids is 
needed to build massive planetary cores, with subsequent accretion of 
large masses of volatiles in the case of gas giants. \citet{dodson} find 
that with 120~$M_{Jup}$ of material (6~MMSN), all four of the Sun's giant 
planets could be made, including non-water ices for speeding up the 
sticking of planet cores. \citet{desch} argues for a steeper-profiled 
initial nebula with later outwards migration of the ice giants, but still 
finds at least of order 60~$M_{Jup}$ (3~MMSN) within 30~AU is 
required\footnote{Some of these authors' assumptions are slightly 
different to ours. \citet{dodson} adopt a gas-to-dust mass ratio of 70 not 
100, and \citet{desch} adopts a smaller MMSN of 13~$M_{Jup}$ for a steep 
disc profile.}. \citet{alibert} have accounted for various properties of 
Jupiter and Saturn in model discs of 35-50~$M_{Jup}$, or 1.75-2.5 MMSN.

According to thse models, giant-planet systems should be rare, if the 
discs seen in the time-snapshot from the millimetre surveys represent true 
budgets of planet-forming material. Only 4~\% of stars host discs of $\ga 
1.75$~MMSN, as required in the most optimistic models for gas giants 
\citep{alibert}. However, the giant planet detection rate is already 7~\% 
in the regime where surveys are close to complete, and extrapolated to 
18~\% out to 20~AU \citep{cumming}, or even higher including lower-mass 
`ice giant' planets \citep{sumi}. Therefore, even if only the most massive 
discs go on to form planets, there is apparently a major discrepancy in 
the statistics, with more exo-planet systems than discs suitable to make them. 

\section{Solutions}

As discussed above, the millimetre dust data may under-estimate the true 
total disc masses, if any of the several assumptions are incorrect. 
Corrections to dust emissivities, relative populations of large and small 
grains, and gas-to-dust proportions could globally shift the distributions 
towards higher masses. However, to reproduce $\approx 18$~\% of 
giant-planet systems \citep{cumming} requires all discs in the 
millimetre surveys of nominally $\ga 10 M_{Jup}$ to be planet-forming, 
whereas the models need at least 35 $M_{Jup}$ of material 
\citep{alibert}. Since an increase of a factor of $\sim 3.5$ to all 
millimetre-derived masses would be both substantial and ad hoc, we now 
consider if other assumptions may be wrong. In particular, we consider 
further solutions to the missing-mass problem, of three kinds:
\begin{enumerate}
\item{mass is used more efficiently than was the case in the Solar System, 
with migrating planets sweeping up much of the disc;}
\item{more mass is `hidden' at small scales in the centre of the disc, 
with no increase of millimetre emission as the dust is optically thick;}
\item{the T Tauri discs observed are only remnants, with most of the dust 
grains growing into larger bodies during the protostellar stage {\bf -- 
this could range from cm-sized rocks up to fully-formed 
planets under the disc instability hypothesis, discussed below}.} 
\end{enumerate}

\subsection{Migration}

The Solar System gas giants are thought to have migrated only slightly 
from their formation positions \citep[e.g.]{minton}, but exo-planets may 
have undergone larger orbit changes. `Hot Jupiters' must have 
migrated significantly inwards, potentially sweeping large disc regions 
clean of material. Highly efficient usage of the disc mass-reservoir -- in 
contrast to $< 10$~\% of the MMSN that ended up in planets \citep{davis} 
-- might hypothetically explain how the discs can be of low-mass and yet 
still form planets. However, long formation times in the outer disc will 
set limits, as cores here would accrete mass slowly and have migrated 
little by the time the disc disperses. Therefore it is implausible that 
the entire disc could be swept up into planets.

If disc-sweeping during migration is important, there should be 
observational signatures in the properties of exo-planets. For example, in 
systems with two planets, the outer one would tend to be of lower mass if 
it has passed through areas already swept clean by the migration of the 
inner body. If the two planets have both ended up close-in to the star, 
this trend should be particularly marked, whereas if the two planets are 
well separated, they could have swept the inner and outer disc 
independently and so not show such a trend. We searched for such 
signatures among multiple-planet systems, using only the parameter space 
$M_{\rm planet} \sin i \geq 0.3 M_{\rm Jup}$ and semi-major axis $a \leq 
3$~AU, where discoveries are thought to be almost complete 
\citep{cumming}. This avoids a bias where lower-mass planets on large 
orbits are harder to detect (i.e. smaller Doppler-wobble) and so such 
systems would be missing from the multi-planet plot. Considering the 
well-studied volume of Sun-like (FGK) stars within 60~pc, 10 systems have 
two planets within our chosen $M,a$ parameter space and $\upsilon$~And has 
three (Figure~3).

\begin{figure}
\label{bindata}
\includegraphics[width=65mm,angle=270]{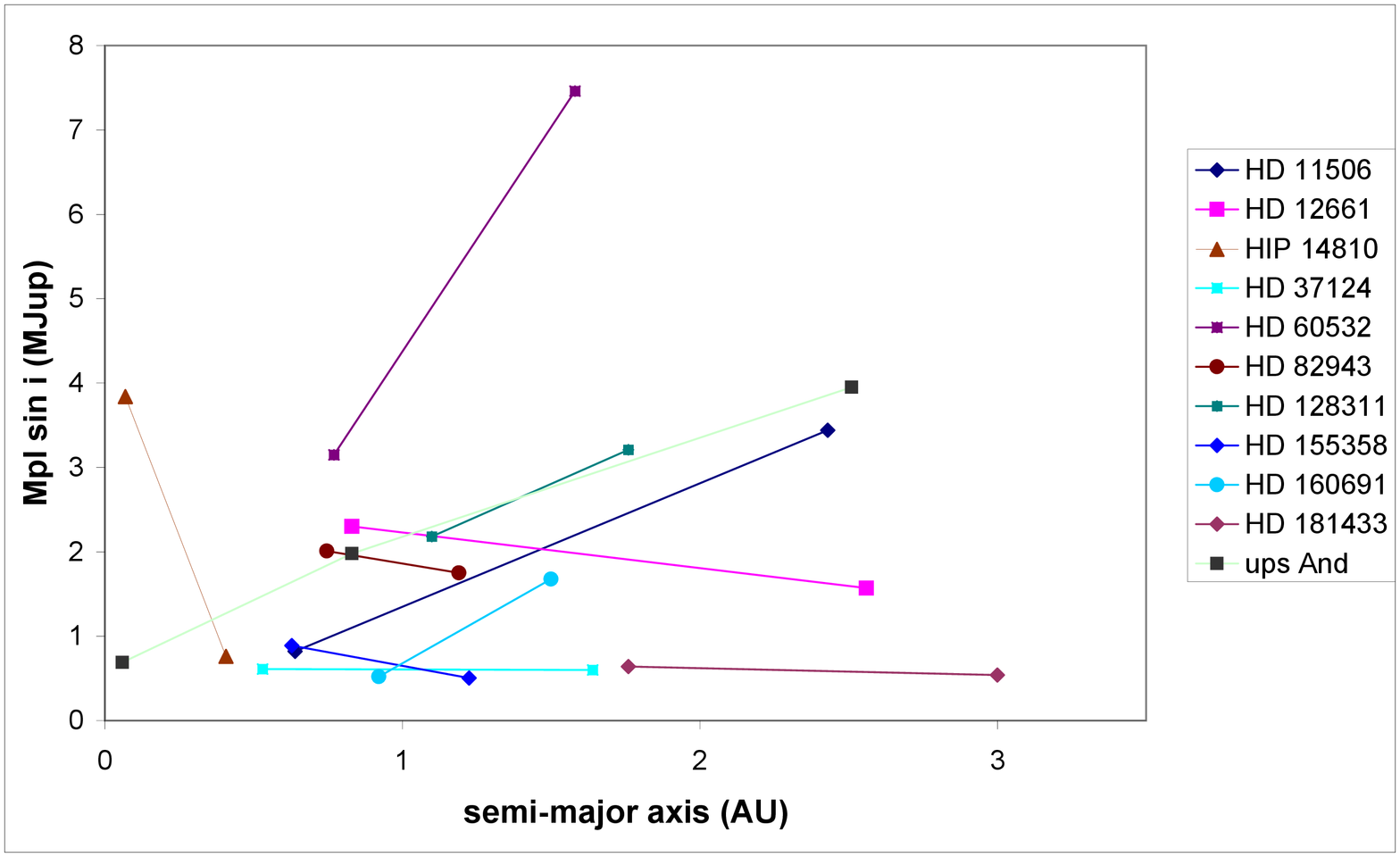}
\caption{Observed masses and semi-major axes for systems with two or three 
planets (from http://exoplanet.eu), in the regime where searches are close 
to complete (see text). Lines connect planets within each system, as 
named at the right. 
}
\end{figure}

These data do not support the idea that accreting planets have swept 
up much of the disc. In particular, only one system in Figure~3 (HIP 
14810) appears to have a `starved' outer planet and both bodies ending up  
near the star. Also, 5/11 of the systems have outer planets that are 
more massive than the inner planet, contradictory to the idea of sweeping 
through already-depleted regions of the disc.

From a theoretical perspective, a further problem with the idea that 
migration can aid planet-building is that most of the gas accretion occurs 
in a runaway phase, with a timescale much shorter than typical migration 
timescales \citep{pollack, ikoma00, bryden00}. Migration is thus 
implausible as a mechanism that allows a planet to sweep up its bulk in 
gas from a large region of the disc. However, migration could aid in the 
growth of the solid planetary core \citep{hourigan,rice03}, and this could 
allow the core to reach a critical mass to attract an atmosphere, while 
there is still sufficient gas in the disc. While migration is thus still 
important to planetary evolution, neither theory or observational 
constraints suggest that it solves the `missing-mass' problem of discs.


\subsection{Centrally-concentrated discs}

Small corrections for opacity are known to be needed when converting 
millimetre dust emission from T Tauri discs into masses. \citet{aw07b} 
estimate that the ratio of optically thick to optically thin submillimetre 
emission is typically around 0.3, as more opacity would result in 
flattening of the spectrum. However, if the discs have a massive central 
inner region, on unresolved scales of tens of AU or less, then this could 
be much more optically thick -- contributing significant mass, but little 
extra millimetre signal. \citet{zhu09,rice09} have recently shown that if 
discs {\it are} massive with respect to the star, then transport of 
angular momentum through disc self-gravity does in fact lead to a pile-up 
of material at smaller radii. A quasi-steady-state is reached in which 
$\sim 80$~\% of the disc mass ends up within 10-20~AU of the star, with a 
drop to lower surface densities in the model outer disc extending to 
50~AU. There is thus a physical basis for the idea of a central mass 
concentration, at scales relevant to planet formation. 	

\begin{figure}
\label{isomass}
\includegraphics[width=75mm,angle=0]{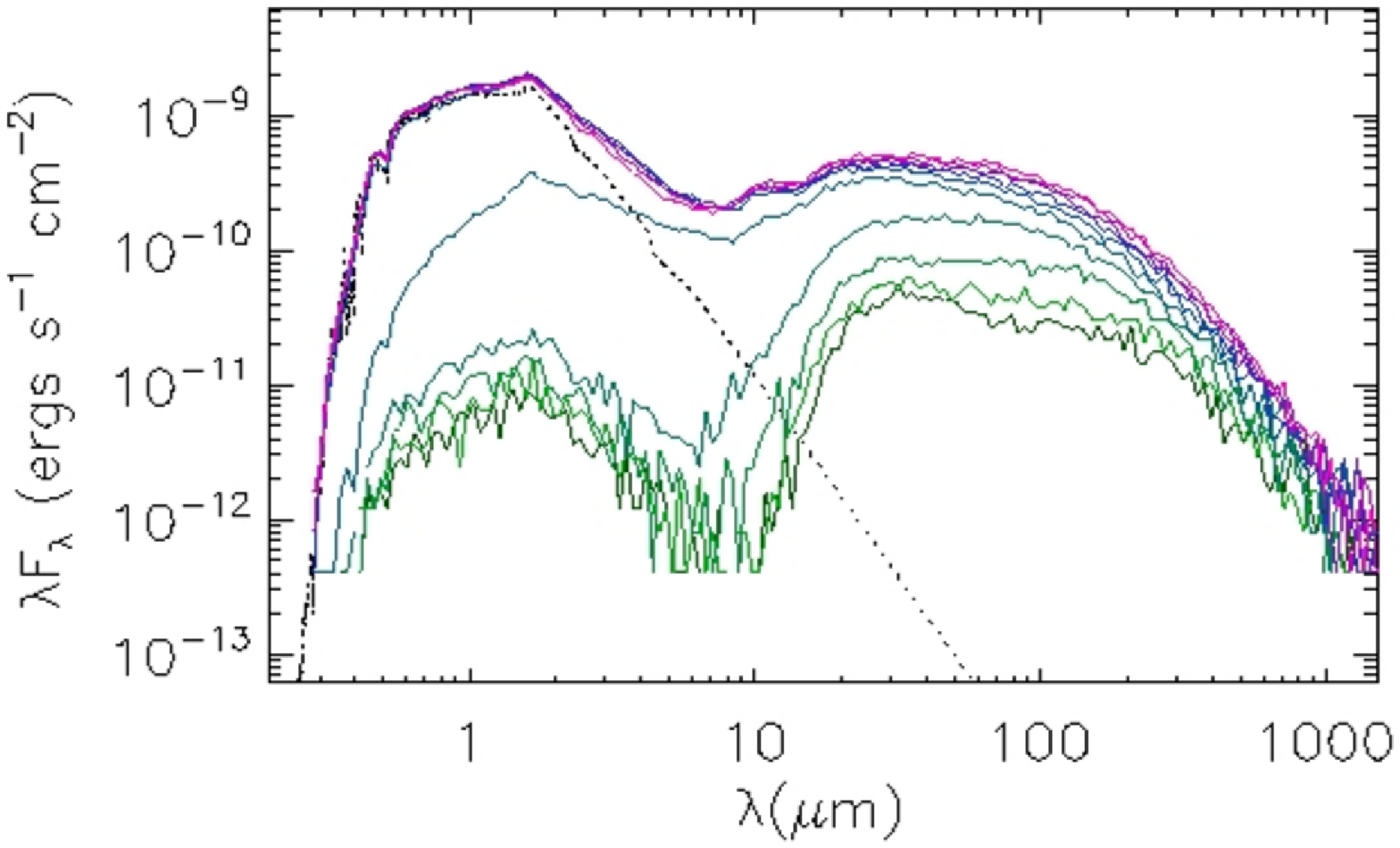}
\includegraphics[width=75mm,angle=0]{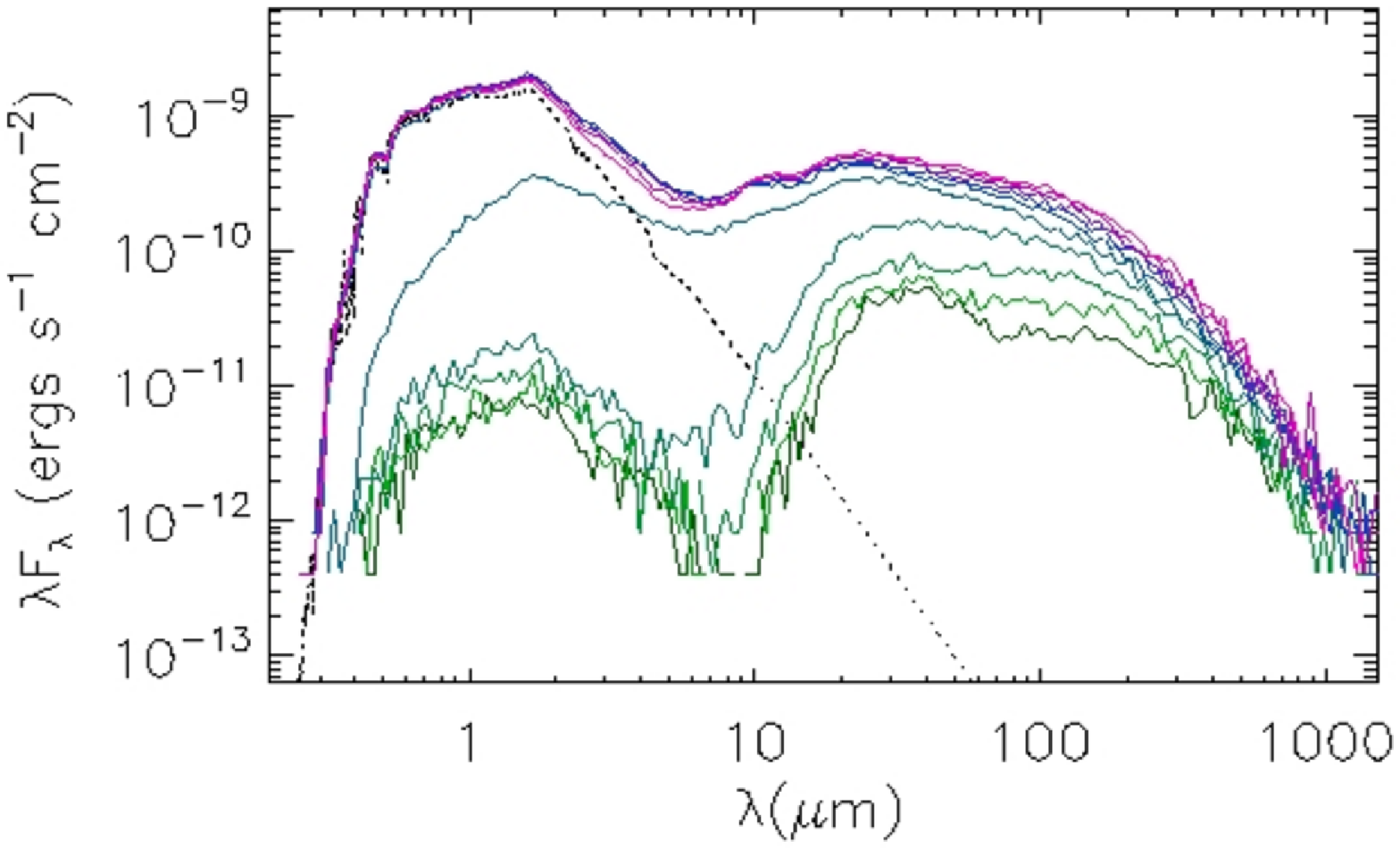}
\caption{Spectral energy distributions for (top) a 0.05~$M_{\odot}$ (50 
Jupiter-mass) disc with a power-law mass distribution, and (bottom) the 
same disc but with more mass added inside 10~AU for a total of 100 Jupiter 
masses. The curves are for different viewing angles, with the stellar 
emission strongly blocked by discs that are closer to edge-on. 
}
\end{figure}

\begin{table*}
 \caption{Properties of Class~0 protostars observed with millimetre 
interferometry. Disc masses are lowest- or single-temperature
results, or from models where a range is given (see papers cited). 
The results shown are for the longest published wavelength, with 
$\ga$ noting that emission around 1~mm may be optically thick. Bolometric 
luminosities are values generally adopted in the literature. The stellar 
masses are derived assuming the protostar's luminosity is powered by 
infall (see text).
}
 \label{tab:protostars}
\begin{tabular}{@{}lcccccl}
  \hline
protostar	& wavelength & $M_{disc}$ & $L_{bol}$	& $M_*(L_{bol})$ 	& $M_{disc}/M_*$ & reference\\
		& (mm)	& ($M_{Jup}$)	 & ($L_{\odot}$)& (M$_{\odot}$)	 	& 		&	    \\
  \hline
L 1448 IRS2	& 2.7		& $\sim$0	& 5		& $\sim$0.08	& $\sim$0	& \citet{kwon}	\\
NGC 1333 IRAS4A	& 2.7		& $\sim$0	& 14-16		& $\sim$0.24	& $\sim$0	& \citet{looney}\\
B 335		& 1.2		& ($\ga$)4	& 3		& $\sim$0.05	& $\ga$0.1	& \citet{harvey}\\
VLA 1623	& 0.8		& ($\ga$)4-10	& 1		& $\sim$0.02	& $\ga$0.4	& \citet{brown}\\
Ser SMM3	& 0.8		& ($\ga$)4-16	& 8		& $\sim$0.13	& $\ga$0.1	& \citet{brown}\\
L 1448 IRS3B	& 2.7		& $\sim$30	& 7		& $\sim$0.11	& $\sim$0.3	& \citet{looney}\\
L 1551 IRS5 S	& 7		& $\sim$30	& $\sim$7	& $\sim$0.11	& $\sim$0.3	& \citet{rod98}	\\
L 483 mm	& 2.6		& $\la$40	& 9		& $\sim$0.14	& $\la$0.3	& \citet{jorgensen}\\
L 1448 C	& 1.4		& ($\ga$)40	& 5		& $\sim$0.08	& $\ga$0.5	& \citet{schoier}\\
Ser SMM1	& 0.8		& ($\ga$)50-80	& 46		& $\sim$0.7	& $\ga$0.1	& \citet{brown}\\
L 1551 IRS5 N	& 7		& $\sim$60	& $\sim$23	& $\sim$0.4	& $\sim$0.2	& \citet{rod98}\\
NGC 1333 IRAS2A	& 2.7		& $\sim$60	& 16-30		& $\sim$0.4	& $\sim$0.2	& \citet{looney}\\
NGC 1333 SVS13B	& 2.7		& $\sim$60	& $\sim$22	& $\sim$0.35	& $\sim$0.2	& \citet{looney}\\
NGC 1333 SVS13A	& 2.7		& $\sim$80	& $\sim$19	& $\sim$0.3	& $\sim$0.3	& \citet{looney}\\
L 1157		& 2.7		& $\sim$110	& 11		& $\sim$0.18	& $\sim$0.6	& \citet{kwon}\\
NGC 1333 IRAS4B	& 2.7		& $\sim$120	& 5		& $\sim$0.08	& $\sim$1.5	& \citet{looney}\\
IRAS 16293 A	& 1.4		& ($\ga$)240	& $\sim$11	& $\sim$0.18	& $\ga$1.4	& \citet{schoier}\\
IRAS 16293 B	& 7		& $\sim$300-400	& 16		& $\sim$0.26	& $\sim$1.4	& \citet{rod05}\\
Ser FIRS1	& 1.1		& ($\ga$)1000	& 46		& $\sim$0.75	& $\ga$1.3	& \citet{enoch} \\
   \hline
 \end{tabular}
\end{table*}

Figure~5 shows representative spectral energy distributions (SED) of the 
star plus disc system -- generated using the HO-CHUNK: 3D radiation 
transfer code \citep{whitney03} -- for a standard power-law distribution 
of disc mass (top panel), and a case where more material has been 
artifically added inside 10~AU to double the total disc material (bottom 
panel). There is very little difference in the SED of the power-law and 
centrally-enhanced discs, especially in the millimetre regime that is 
canonically `mass tracing'.  In fact, the centrally condensed disc has a 
submillimetre flux that is slightly lower than that from 
the lower-mass, power-law disc.  Therefore, if real discs are in fact 
massive, a central pile-up would be both theoretically predicted 
\citep{rice09} and not detected in millimetre images where the inner disc 
is unresolved. This provides a potential solution to the missing mass, at 
least for discs with moderately high mass estimates already.

\subsection{Disc masses at early times}

As discussed above, discs have higher mass at earlier evolutionary stages. 
At Class~I, the fraction of objects hosting 2-25~MMSN is around 18~\%, in 
the regime of Jupiter-building models, and also matching the 18~\% of 
gas-giant planet systems inferred around mature stars \citep{cumming}. 
However, if planetary cores can grow this early on -- which could indeed 
occur if migration were to accelerate the core growth as noted above -- 
then in principle the process could even start in the Class~0 protostar 
phase, within the first 0.1~Myr. At this time the central object is only 
partly accreted, and much of the mass of the system is still in a 
circumstellar envelope. Millimetre photometry will be dominated by the 
envelope flux, but many of the Class~0 objects have now been studied with 
millimetre-wavelength interferometry. This resolves out the large-scale 
envelope emission, leaving in most cases a compact or point-like signal 
attributed to a very young disc.

Table~3 lists these interferometry-based results, in order of increasing 
disc mass. The mass estimates were made with different methods and assumed 
temperatures and emissivities, and may be lower limits at $\approx 1$~mm 
if the inner disc has non-negligible opacity, so only rough comparisons 
may be made. However, among these 19 Class~0 objects, half now host 
upwards of 2~MMSN, increased three-fold over the equivalent population 
within Class~I. This continues the trend of finding more substantial discs 
at each earlier phase studied, and is very promising for forming many 
future planetary systems. {\bf If these larger dust masses at the Class~0 
stage represent most closely a true `initial mass reservoir' for planet 
formation, then the small quantities observed during Classes I, FS and II 
suggest that dust has been converted into larger pebbles and boulders, and 
perhaps planetesimals if core accretion has proceded successfully. In the 
grain growth models of \citet{dd05}, metre-sized boulders are abundant by 
ages of a few $10^5$ years, i.e. in the Class I stage, and planetesimals 
can aggregate into planetary cores over a similar timescale 
\citep{hubickyj}. }

\subsubsection{Planet formation by gravitational instability}

In addition, as the protostar is largely assembled in the Class~0 phase, 
it must for part of this period have much less than its final mass. 
Therefore, as the discs are mostly substantial, $M_{disc}/M_{star}$ can be 
high, raising the possibility of forming planets by the alternative 
mechanism where gravitational instability leads to disc fragmentation. To 
attempt to quantify this, we estimate masses for the protostars assuming 
that the bolometric luminosity is supplied by the energy released in 
infall. The accretion rate is taken to be $10^{-5}$~M$_{\odot}$~yr$^{-1}$, 
i,e, building most of an $\sim 1~M_{\odot}$ star in about 0.1~Myr 
\citep{evans}. The mass of the protostar $M_{star}$ is then derived from 
$L_{bol} = (G M_{star} dM/dt) / R_{star}$, taking $R_{star}$ to be 
$\approx 5 R_{\odot}$ after \citet{am}. The protostellar radii have not 
been measured but for pre-main-sequence phases should be moderately 
independent of mass \citep{whitney04}. $M_{disc}/M_{star}$ values may 
be uncertain by factors up to a few, as disc opacities, stellar 
radii and accretion rates are not known for individual systems.
 
The estimates in Table~3 suggest that half of the discs should be unstable 
to fragmentation, with $M_{disc}/M_{star}$ of roughly 0.3 or more. In four 
cases, the disc appears to actually out-weigh the star, as in some models 
of \citet{hg05,rma}. The stellar masses are low, as expected if the stars 
are not yet fully built. The protostars estimated to be the most massive 
also tend to have higher-mass discs, as do the objects in binary systems 
(i.e. names ending in A, B, N or S). However, most of the short-wavelength 
data happens to be for the single systems, so these could simply have disc 
masses that are under-estimated due to dust opacity.

One problem with the idea of forming planets early on via disc 
fragmentation is that theoretical calculations, both semi-analytic 
\citep{matzner05,rafikov05, rice09,clarke09} and numerical 
\citep{boley06,stamatellos08}, suggest that planetary mass bodies will not 
form inside $\sim 50$~AU. It is likely that these extremely massive 
Class~0 discs would instead become globally unstable \citep{lodato05}, 
moving large amounts of mass to large radii, where fragmentation could 
produce substellar, or even stellar, mass companions 
\citep{stamatellos09}. However, the presence of spiral structures in the 
massive disc could help to collect together solid particles 
\citep{haghighipour03, rice04} and thus accelerate planet formation via 
the mechanism of core accretion. Thus massive discs at very early times 
may still be a promising signpost to abundant planetary systems around 
mature main-sequence stars.
 
\section{Conclusions}

The compilation of millimetre data for circumstellar discs in 1-2~Myr-old 
star formation regions confirms that the {\bf observable} mass reservoirs 
are small, at these snapshots in time. The fraction of discs {\bf 
apparently} capable of forming gas giants by core accretion is generically 
less than the fraction of observed exo-planet systems. An early start to 
planetary growth is favoured, as the fractions of exo-planets and suitable 
proto-planetary discs do match in the Class~I stage. We also considered 
other solutions to the `missing mass' problem. Sweeping up the disc 
material efficiently into planets has both observational and theoretical 
difficulties. More promisingly, viscous evolution through disc 
self-gravity could produce a central concentration of mass that would be 
optically thick even in the millimetre, and so essentially invisible. At 
very early times (Class~0), the discs are both massive and at a high 
mass-fraction with respect to the star, so growth into larger bodies could 
even start at this very early phase when the star is only part-completed. 
In conclusion, most of the {\bf dust discs seen around Class~II classical 
T~Tauri stars are too insubstantial for commencing to grow planets, and so 
these should be relics, in the sense that the initial mass reservoir in 
dust grains has already been converted to larger forms with little 
millimetre emission. This growth process around a particular star could 
have reached anywhere from small rocks up to planetesimals or even 
completed gas giant planets, especially if the rapid disc instability 
mechanism operates.}

\section*{Acknowledgments} 

We thank STFC and SUPA for support, and Kenny Wood for insightful remarks 
that inspired the early stages of this project.

\label{lastpage}


\begin{thebibliography}{99}
\bibitem[\protect\citeauthoryear{Alibert et al.}{2005}]{alibert}
Alibert Y., Mousis O., Mordasini C., Benz W., 2005, ApJ 626, L57
\bibitem[\protect\citeauthoryear{Andr\'{e} \& Montmerle}{1994}]{am}
Andr\'{e} P., Montmerle T., 1994, ApJ 420, 837
\bibitem[\protect\citeauthoryear{Andrews \& Williams}{2007b}]{aw07b}
Andrews S.A., Williams J.P., 2007, ApJ 659, 705
\bibitem[\protect\citeauthoryear{Andrews \& Williams}{2007a}]{aw07}
Andrews S.A., Williams J.P., 2007, ApJ 671, 1800 
\bibitem[\protect\citeauthoryear{Andrews \& Williams}{2005}]{aw05}
Andrews S.A., Williams J.P., 2005, ApJ 631, 1134
\bibitem[\protect\citeauthoryear{Bary et al.}{2008}]{bary}
Bary J.S., Weintraub D.A., Shukla S.J., Leisenring J.M., Kastner J.H., 
2008, ApJ 678, 1088
\bibitem[\protect\citeauthoryear{Boley et al.}{2006}]{boley06}
Boley A.C., Mejia A.C., Durisen R.H., Cai K., Pickett M.K., D'Alessio
P., 2006, ApJ, 651, 517
\bibitem[\protect\citeauthoryear{Brown et al.}{2000}]{brown}
Brown D.W., Chandler C.J., Carlstrom J.E., Hills R.E., Lay O.P., 
Matthews B.C., Ricer J.S., Wilson C.D., 2000, MNRAS 319, 154
\bibitem[\protect\citeauthoryear{Bryden et al.}{2000}]{bryden00}
Bryden G., Lin D.N.C., Ida S., 2000, ApJ, 544, 481
\bibitem[\protect\citeauthoryear{Carpenter}{2002}]{carpenter} 
Carpenter J.M., 2002, AJ 124, 1593
\bibitem[\protect\citeauthoryear{Clarke}{2009}]{clarke09}
Clarke C.J., 2009, MNRAS, 396, 1066
\bibitem[\protect\citeauthoryear{Cumming et al.}{2008}]{cumming}
Cumming A., Butler R.P., Marcy G.W., Vogt S.S., Wright J.T.,
Fischer D.A., 2008, PASP 120, 531
\bibitem[\protect\citeauthoryear{Davis}{2005}]{davis} 
Davis S. S., 2005, ApJ 627, L153
\bibitem[\protect\citeauthoryear{Dent et al.}{2005}]{dent} 
Dent W.R.F., Greaves J.S., Coulson I M., 2005, MNRAS 359, 663
\bibitem[\protect\citeauthoryear{Desch}{2007}]{desch}
Desch S.J., 2007, ApJ 671, 878
\bibitem[\protect\citeauthoryear{Dodson-Robinson et al.}{2009}]{dodson}
Dodson-Robinson S.E., Willacy K., Bodenheimer P., Turner N.J.,  
Beichman C.A., 2009, Icarus 200, 672
\bibitem[\protect\citeauthoryear{Draine}{2006}]{draine} Draine
Draine B.T., 2006, ApJ 636, 1114
\bibitem[\protect\citeauthoryear{Dullemond \& Dominik}{2005}]{dd05}
Dullemond C.P., Dominik C., 2005, A\&A 434, 971
\bibitem[\protect\citeauthoryear{Enoch et al.}{2009}]{enoch}
Enoch M.L., Corder S., Dunham M.M., Duch\^{e}ne G., 2009, ApJ 707, 103
\bibitem[\protect\citeauthoryear{Ercolano et al.}{2009}]{ercolano}
Ercolano B., Clarke C.J., Drake J.J., 2009, ApJ 699, 1639
\bibitem[\protect\citeauthoryear{Evans et al.}{2009}]{evans}
Evans N.J. et al., 2009, ApJS 181, 321
\bibitem[\protect\citeauthoryear{Haghighipour \& Boss}{2003}]{haghighipour03}
Haghighipour N., Boss A.P., 2003, ApJ, 583, 996
\bibitem[\protect\citeauthoryear{Harvey et al.}{2003}]{harvey} 
Harvey D.W.A., Wilner D.J., Myers P.C., Tafalla M., Mardones D., 2003, 
ApJ 583, 809
\bibitem[\protect\citeauthoryear{Henning et al.}{1993}]{henning}
Henning T., Pfau W., Zinnecker H., Prusti T., 1993, A\&A 276, 129
\bibitem[\protect\citeauthoryear{Hourigan \& Ward}{1984}]{hourigan}
Hourigan K., Ward W.R. 1984, Icarus, 60, 29
\bibitem[\protect\citeauthoryear{Hubickyj et al.}{2005}]{hubickyj}
Hubickyj O., Bodenheimer P., Lissauer J. J., 2005, Icarus 179, 415
\bibitem[\protect\citeauthoryear{Hueso \& Guillot}{2005}]{hg05}
Hueso R., Guillot T., 2005, A\&A 442, 703
\bibitem[\protect\citeauthoryear{Ikoma et al.}{2000}]{ikoma00}
Ikoma M., Nakazawa K., Emori H., 2000, ApJ, 537, 1013
\bibitem[\protect\citeauthoryear{J\o rgensen et al.}{2004}]{jorgensen}
J\o rgensen J.K., 2004, A\&A 424, 589
\bibitem[\protect\citeauthoryear{Kenyon \& Hartmann}{1995}]{kenyon}
Kenyon S.J., Hartmann L., 1995, ApJS 101, 117
\bibitem[\protect\citeauthoryear{Kwon et al.}{2009}]{kwon}
Kwon W., Looney L.W., Mundy L.G., Chiang H.-F., Kemball A.J., 2009, AJ 
696, 841
\bibitem[\protect\citeauthoryear{Lavalley et al.}{1992}]{lavalley}
Lavalley M., Isobe T., Feigelson E., 1992, A.S.P. Conference Series, 
Vol. 25, p. 245
\bibitem[\protect\citeauthoryear{Lodato \& Rice}{2005}]{lodato05}
Lodato G., Rice W.K.M., 2005, MNRAS, 358, 1459
\bibitem[\protect\citeauthoryear{Looney et al.}{2003}]{looney} 
Looney L.W., Raab W., Poglitsch A., Geis N., 2003, ApJ 597, 628
\bibitem[\protect\citeauthoryear{Mann \& Williams}{2009}]{mann}
Mann R.K., Williams J.P., 2009, ApJ 699, L55
\bibitem[\protect\citeauthoryear{Matzner \& Levin}{2005}]{matzner05}
Matzner C.D., Levin Y., 2005, ApJ, 628, 817
\bibitem[\protect\citeauthoryear{Mer\'{i}n et al.}{2008}]{merin}
Mer\'{i}n B. et al., 2008, ApJS 177, 551
\bibitem[\protect\citeauthoryear{Minton \& Malhotra}{2009}]{minton}
Minton D.A., Malhotra R., 2009, Nature 457, 1109
\bibitem[\protect\citeauthoryear{Muench et al.}{2007}]{muench}
Muench A.A., Lada C.J., Luhman K.L., Muzerolle J., Young E., 2007, AJ 134, 
411
\bibitem[\protect\citeauthoryear{N\"{u}rnberger et al.}{1997}]{nurnberger}
N\"{u}rnberger D., Chini R., Zinnecker H., 1997, A\&A 324, 1036
\bibitem[\protect\citeauthoryear{Palla \& Stahler}{2000}]{palla} 
Palla F., Stahler S.W., 2000, ApJ 540, 255
\bibitem[\protect\citeauthoryear{Pollack et al.}{1996}]{pollack}
Pollack J.B., Hubickyj O., Bodenheimer P., Lissauer J.J., Podolak M.,
Greenzweig Y., 1996, Icarus 124, 62
\bibitem[\protect\citeauthoryear{Porras et al.}{2003}]{porras} 
Porras A., Christopher M., Allen L., Di Francesco J., Megeath S.T., 
Myers P.C., 2003, AJ 126, 1916
\bibitem[\protect\citeauthoryear{Rafikov}{2005}]{rafikov05}
Rafikov R.R., ApJ, 621, L69
\bibitem[\protect\citeauthoryear{Rice et al.}{2009}]{rma}
Rice W.K.M., Mayo J.H., Armitage P.J., 2009, MNRAS (online early, Dec.)
\bibitem[\protect\citeauthoryear{Rice \& Armitage}{2009}]{rice09}
Rice W.K.M., Armitage P.J., 2009, MNRAS, 396, 2228
\bibitem[\protect\citeauthoryear{Rice et al.}{2004}]{rice04}
Rice W.K.M., Lodato G., Pringle J., Armitage P.J., Bonnell I.A., 2004, 
MNRAS, 372, L9
\bibitem[\protect\citeauthoryear{Rice et al.}{2003}]{rice03}
Rice W.K.M., Armitage P.J., Bate M.R., Bonnell I.A. 2003, MNRAS, 346, L36
\bibitem[\protect\citeauthoryear{Rodmann et al.}{2006}]{rodmann} 
Rodmann J., Henning Th., Chandler C.J., Mundy L.G., Wilner D.J.,
2006, A\&A 446, 211
\bibitem[\protect\citeauthoryear{Rodr\'{i}guez et al.}{2005}]{rod05}
Rodr\'{i}guez L.F., Loinard L., D'Alessio P., Wilner D.J., Ho P.T.P, 
2005, ApJ 621, L133
\bibitem[\protect\citeauthoryear{Rodr\'{i}guez et al.}{1998}]{rod98}
Rodr\'{i}guez L.F. et al., 1998, Nature 395, 355
\bibitem[\protect\citeauthoryear{Sch\"{o}ier et al.}{2004}]{schoier}
Sch\"{o}ier F.L., J\o rgensen J.K., van Dishoeck E.F., Blake G.A., 2004, 
A\&A 418, 185
\bibitem[\protect\citeauthoryear{Stamatellos \& Whitworth}{2009}]{stamatellos09}
Stamatellos D., Whitworth A.P., 2009, MNRAS 392, 413
\bibitem[\protect\citeauthoryear{Stamatellos \& Whitworth}{2008}]{stamatellos08}
Stamatellos D., Whitworth A.P., 2008, A\&A, 480, 879
\bibitem[\protect\citeauthoryear{Sumi et al.}{2010}]{sumi}
Sumi T. et al., 2010, ApJ 710, 1641
\bibitem[\protect\citeauthoryear{Ward-Thompson et al.}{2007}]{gould} 
Ward-Thompson D. et al., 2007, PASP 119, 855
\bibitem[\protect\citeauthoryear{Whitney et al.}{2004}]{whitney04}
Whitney B.A., Indebetouw R., Bjorkman J.E., Wood K., 2004, ApJ 617, 1177
\bibitem[\protect\citeauthoryear{Whitney et al.}{2003}]{whitney03}
Whitney B.A., Wood K., Bjorkman J.E., Wolff M.J., 2003, ApJ, 591, 1049
\bibitem[\protect\citeauthoryear{Wright et al.}{2009}]{wright}
Wright J.T., Upadhyay S., Marcy G.W., Fischer D.A., Ford E.B.,  
Johnson J.A., 2009, ApJ 693, 1084
\bibitem[\protect\citeauthoryear{Wyatt et al.}{2007}]{wyatt}
Wyatt M.C., Clarke C.J., Greaves J.S., 2007, MNRAS 380, 1737
\bibitem[\protect\citeauthoryear{Zhu et al.}{2009}]{zhu09}
Zhu Z., Hartmann L., Gammie C.F., 2009, ApJ, 694, 1045
\end{thebibliography}
\end{document}